\documentclass[draft,twocolumn,prb,amssymb,showpacs]{revtex4}
\input epsf
\begin{document}

\title{Weak localization in macroscopically inhomogeneous
two-dimensional \\ systems: a simulation approach}
\author{A.~V.~Germanenko}
\email{Alexander.Germanenko@usu.ru}
\author{G.~M.~Minkov}
\author{O.~E.~Rut}
\affiliation{Institute of Physics and Applied Mathematics, Ural
State University, 620083 Ekaterinburg, Russia}

\date{\today}

\begin{abstract}
A weak-localization effect has been studied in macroscopically
inhomogeneous 2D system. It is shown, that although the real phase
breaking length tends to infinity when the temperature tends to
zero, such a system can reveal a saturated behavior of the
temperature dependence of that parameter, which is obtained from
the standard analysis of the negative magnetoresistance and
usually identified by experimentalists with the phase braking length.
\end{abstract}
\maketitle

The numerous experimental as well as theoretical papers are
devoted to the weak localization in semiconductor structures and
metal films. The problem was most intensively studied in the
eightieth years. The theory of weak localization, which adequately
describes experimental data, was developed. It gave  simple
analytical expressions for quantum correction to conductivity
which allowed to determine a phase breaking time in real electron
systems experimentally.  The comprehensive review of the status of
the problem at that time is given in Ref.~\onlinecite{c1,c2}.

In recent years the interest to this problem reappeared. One of
the reason is that the new experimental results have been
obtained. One of them is the saturation of the temperature
dependence of the phase breaking time $\tau_\varphi$. \cite{satT1}
These papers cause a storm discussion in the literature (see,
e.g., Refs.\ \onlinecite{satT3,satT5,satT6} and references
therein) and stimulate a new flux of the papers concerning the
weak localization.

A standard fitting procedure is used to analyze the experimental
data and determine the phase breaking time practically in all
cases: experimental magnetic-field dependencies of
magnetoresistance are fitted to the theoretical
ones,\cite{hik,schm} the phase breaking time is the fitting
parameter.

Another approach to examination of the negative magnetoresistance
due to weak localization has been presented in Ref.\
\onlinecite{our1}. It is based on a quasi-classical treatment of
the problem\cite{chak,dyak,dmit} and an analysis of the statistics
of closed paths of a classic particle moving with scattering over
2D plane. This method has been applied to study the weak
localization in InGaAs/GaAs heterostructures with
single\cite{our2} and double\cite{our3} quantum wells.

The foregoing relates to homogeneous systems revealing only weak
disorder on microscopic length scale. Of special interest are
inhomogeneous systems, e.g., granular CuO films\cite{AronAED} and
percolating gold films.\cite{Pal,Dump,fried} For interpretation of
the experimental results the concept of anomalous diffusion is
widely used. It should be mentioned that the transition from the
weak localization to strong localization regime, when the system
parameters or external conditions are varied, can also cross the
macroscopically inhomogeneous state in originally homogeneous 2D
systems.

In the present paper we analyze the weak-localization negative
magnetoresistance in macroscopically inhomogeneous systems using
approach developed in Ref.\ \onlinecite{our1}. The purpose is to
show how the parameters, determined from the magnetoresistance
experiments, match their real values in the systems with different
inhomogeneity. Moreover, we show that this approach allows
experimentally to distinguish what geometry of inhomogeneity is
dominant in the system under investigation. As in Ref.\
\onlinecite{our1} the results presented have been obtained from
the numerical simulation of classical motion of particle.

\begin{figure}
\epsfclipon
 \epsfxsize=\linewidth
 \vspace{1cm}
 \epsfbox{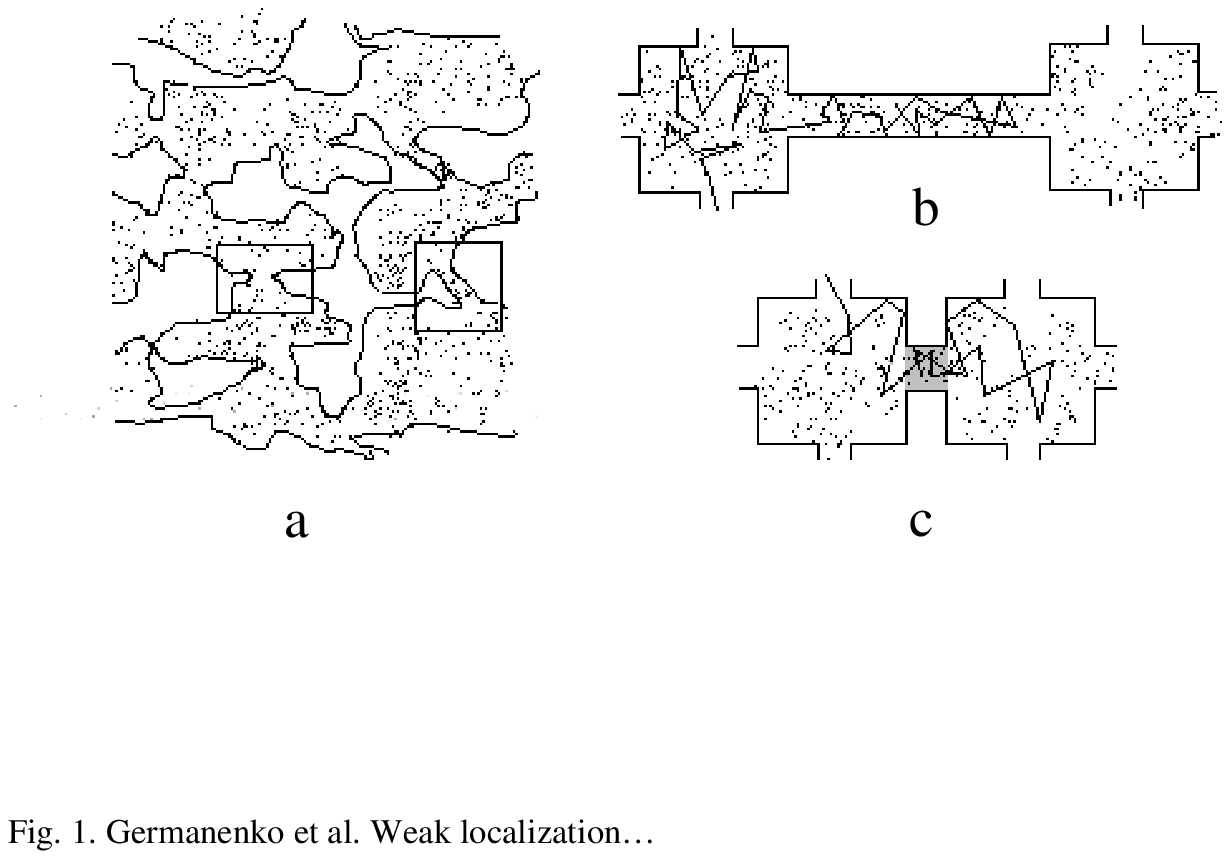}
 \caption{(a)~Sketch of inhomogeneous 2D system. The key parts
determining the conduction are enclosed in rectangles.
Non-conducting areas are white. Points represent scatterers. (b),
(c)~The models of long and square channels, respectively, used in
simulation. Polygonal lines show particle trajectories. Shadowing
in (c) shows the area with smaller value of the Fermi
quasimomentum.} \label{fig1}
\end{figure}

We suppose that a macroscopically inhomogeneous 2D system consists
of a number of puddles which are connected one with other by means
of channels (Fig.~\ref{fig1}(a)). The transport through the
puddles and channels is diffusive, i.e. their dimensions are much
greater than the mean free path of electrons. In this case the
quasi-classical approach to consideration of the problem can be
applied. We focus our attention on the case when the system is in
inhomogeneous regime, i.e, when $L_\varphi<\xi_p$,\cite{Stauf}
where $\xi_p$ is the percolation correlation length,
$L_\varphi=\sqrt{D\tau_\varphi}$, the diffusion constant is given
by $D=v_F^2\tau/2$ with $v_F$ as the Fermi velocity and $\tau$
representing the transport elastic mean-free time. Furthermore,
we have idealized the situation and supposed that all the
channels and puddles are identical, and our model system is a
series-parallel connection of the elements, each of them is a
channel and two half-puddles connected (Fig.~\ref{fig1}(b,c)). We
reduce thus the problem on the infinite inhomogeneous 2D system to
the problem on the weak localization in one element, which is a
diffusive constriction connecting two diffusive puddles. Because
the channel and two paddles are connected in series, the
correction to the conductance of the element is
\begin{equation}
\delta\sigma=\frac{1}{R_c+R_p}\left(\frac{R_c}{R_c+R_p}
\frac{\delta\sigma_c}{\sigma_c}+
\frac{2R_p}{R_c+R_p}\frac{\delta\sigma_p}{\sigma_p}\right),
\label{eq07}
\end{equation}
where $R_c$, $R_p$ are the resistances, $\sigma_{c}$, $\sigma_{p}$
and $\delta\sigma_{c}$, $\delta\sigma_{p}$ are the conductivity
and correction to the conductivity, respectively. Indexes $c$ and
$p$, hereafter, refer to the channel and puddle, respectively.

The quantities $\delta\sigma_c$ and $\delta\sigma_p$ are
determined by the statistics of closed paths and can be found as
follows \cite{chak,dyak,dmit}
\begin{equation}
 \delta\sigma_{c,p}=-2\pi l_{c,p}^2 G_0 {\cal W}_{c,p}
 \label{eq08}
\end{equation}
where $l$ denotes the mean-free path within the channel ($l_p$) or
puddle ($l_p$), $G_0=e^2/(2\pi^2 \hbar)$, ${\cal W}$ stands for
the classical quasi-probability density for an electron to return
to the start point within the channel (${\cal W}_c$) or puddle
(${\cal W}_p$). With $\sigma=\pi k_F l\, G_0$, where $k_F$ is the
Fermi quasimomentum, we find
\begin{equation}
\delta\sigma=\frac{2}{\pi G_0
(R_c+R_p)^2}\left(\frac{K_c}{(k_F^c)^2} {\cal W}_c+
2\frac{K_p}{(k_F^p)^2}{\cal W}_p \right),
 \label{eq09}
\end{equation}
where $K_c$ and $K_p$ are the length to width ratio of channel and
puddle, respectively. Using formalism of Ref.~\onlinecite{our1},
we can write for qusi-probabilities ${\cal W}_c$, ${\cal W}_p$ in
the presence of external magnetic filed and inelastic scattering
processes, destroying the phase coherence:
\begin{equation}
    {\cal W}_{i}(B)= \int_{-\infty}^\infty dS\ W_i(S)
\exp\left(-\frac{\overline{L}_i}{l_\varphi^{i}}\right)\cos\left(
  \frac{2\pi B S}{\Phi_0}\right),
\label{eq1}
\end{equation}
where $W_i(S)$ are the area distribution functions of closed
paths, starting within the channel ($i=c$) or puddle ($i=p$),
defined in such a way that $W_i(S)dS$ gives the classical
probability density of return to the starting point with area
enclosed $S$ from the interval $(S,S+dS)$,
$l_\varphi^i=v_F^i\tau_\varphi^i$, $\overline{L}_i$ is the
average length of closed paths with a given area, calculated by
appropriate manner [Eq.\ (6) in Ref.\ \onlinecite{our1}],
$\Phi_0$ is the elementary flux quantum. Expressions (\ref{eq09})
and (\ref{eq1}) allow to obtain the magnetic field and
temperature (through $l_\varphi^i$) dependence of conductance of
our inhomogeneous model system.

Thus, the relative contributions of channel and puddles to the
interference correction are determined not only by the statistics
of closed paths, but geometry of channel and puddles and the
values of the Fermi quasimomentum in them.

In the present paper we consider the situation when the channel,
which connects the square puddles ($K_p=1$), mainly determines the
conductance of whole 2D system and interference correction to it
in actual range of parameters, i.e., the first term in
Eq.~(\ref{eq09}) is greater than the second one. Two opposite
cases are analyzed: (i) $k_F^c=k_F^p$, $K_c\gg 1$ (long channel);
(ii) $k_F^c\ll k_F^p$, $K_c=1$ (square channel). In the second
case the lower value of the quasimomentum within the channel
leads to arising of quasimomentum dependent reflection of
electrons from the opened ends of channel. Those electrons, which
attempt to enter the channel at an acute angle to the
channel-puddle border ($\theta<\theta'$), are reflected and
return to the puddle. All the electrons escaping the channel do
this freely without any reflection. The value of $\theta'$ is
determined by $k_F^c$ to $k_F^p$ ratio. The reflection can be
specular, if the border is smooth, or diffusive, if the border is
ragged. For simplicity we suppose that the border is straight
line and reflection is specular.

Let us describe the simulation procedure.  The element is
represented as a lattice which contour corresponds to the geometry
of element. The scatterers with given cross-section are placed in
a part of lattice sites with the use of a random number generator.
A particle is launched from some random point, then it moves with
a constant velocity along straight lines, which happen to be
terminated by collisions with the scatterers. After collision it
changes the motion direction. If the particle collides with the
walls of element, it is specularly reflected. If the particle
passes near the starting point at the distance less than $d/2$
(where $d$ is a prescribed value, which is small enough), the
path is perceived as being closed. Its length and enclosed
algebraic area are calculated and kept in memory. The particle
walks over the element until it reaches one of the channel
belonging to another element. As this happens one believes that
the particle has left to infinity and will not return. A new
start point is chosen and all is repeated.

The parameters used in simulation are the following: $d=5$
[hereafter we measure length and area in units of lattice
parameter and its square, respectively]; the number of launches
$I_s=10^5..10^6$; the length and width of channel are $4000$ and
$400$ for long channel, $400$ and $400$ for square channel. The
dimension of each puddle is $6000\times 6000$ for all cases. The
scattering is isotropic, the  cross-section of the scatterers is
equal to 7. The density of scatterers is such that the mean-free
path is about $40$ for both cases.  For the case of square channel
we use $(k_F^p/k_F^c)^2=10$, that gives the value of $\theta'$
about 70 degrees. For illustration, let us set the lattice
parameter equal to $5$ \AA. Then our model provides an example of
2D system with local concentration of scatterers $1.5\times
10^{12}$ cm$^{-2}$, $l=200$ \AA\  and $B_{tr}\simeq 0.8$~T, where
$B_{tr}=\hbar c/(2el^2)$ is so called transport magnetic field.

\begin{figure}
\epsfclipon
 \epsfxsize=\linewidth
 \epsfbox{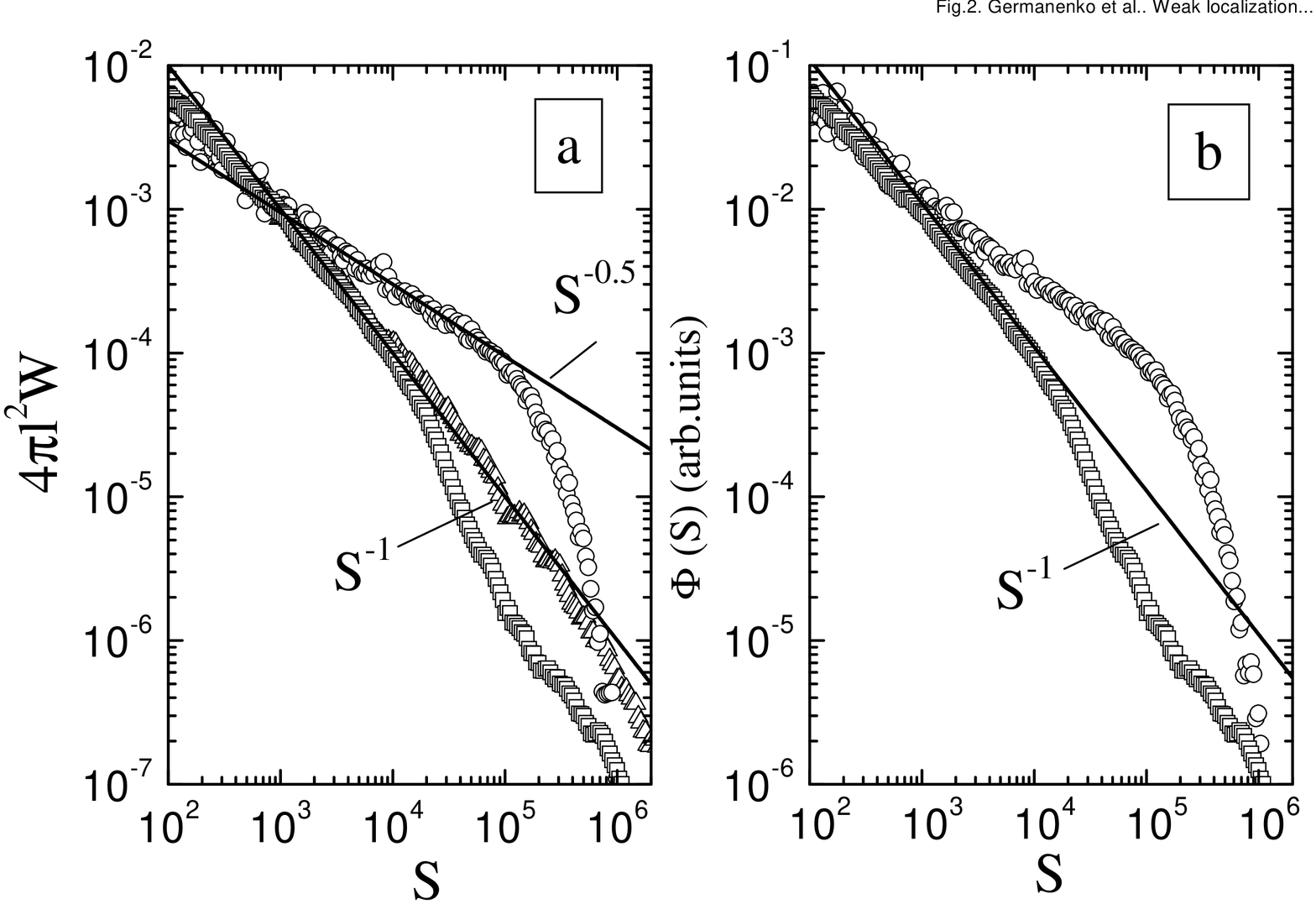}
 \caption{(a) The area distribution functions $W_i(S)$ (a) and $\Phi(S)$
function (b). Symbols are the simulation results for the case of
long ($\circ$), square ($\square$) channels, and for the paths
starting within the puddle ($\vartriangle$).  Only the positive
area range is presented. Results obtained for negative algebraic
areas are identical.} \label{fig2}
\end{figure}

We first turn to the area distribution function $W(S)$
(Fig.~\ref{fig2}). The area range we are interested in is
$S\gtrsim l^2$, because just the long closed paths determine the
low field negative magnetoresistance at low temperatures. As is
seen, $4\pi l^2 W_p(S)$ mostly follows $S^{-1}$ dependence. This
is in agreement with the result of diffusion theory for infinite
2D system,\cite{our1,samokh} which gives $4\pi l^2
W(S)=S^{-1}\tanh(\pi S/l^2)\simeq S^{-1}$ for $S>l^2$.

For the channels, the dependence $4\pi l^2 W(S)$ looks more
complicated. For the long channels it is close to $S^{-0.5}$-law
within the area range $1\times 10^3-5\times 10^4$. It is precisely
this behavior that is theoretically predicted for diffusive motion
over infinitely long, narrow strip. A particular interest is
drastic decrease of the area distribution function evident for
$S>10^5$. The origin is that the particle, moving along a long
trajectory, reaches one of the ends of long channel, escapes the
channel, and carries on the motion mainly within the puddles.

As for square channel, the $4\pi l^2 W_c$-versus-$S$ plot shows
the behavior close to $S^{-1}$ up to $S\simeq 10^4$, and reveals
strong decrease at large areas as in previous case. This decrease
is caused by the fact that a particle crosses the channel/puddle
border freely only in one direction: when it moves from the
channel to puddle. It cannot easily return back due to reflection
from the opened ends of channel resulting from the difference of
the Fermi quasimomentum in channel and puddle discussed above.

In contrast to the area distribution functions, the dependencies
$\overline{L}(S)$ are similar for all three cases.

In Figure \ref{fig2} the function
\begin{equation}
\Phi(S)=\frac{K_c}{(k_F^c)^2} W_c(S)+
2\frac{K_p}{(k_F^p)^2}W_p(S),
 \label{eq11}
\end{equation}
which determines the magnetic field dependence of interference
quantum correction of the element, is plotted. Namely this
function rather than the area distribution function as in
homogeneous 2D case will be obtained when the approach to
analysis of the negative magnetoresistance suggested in
Refs.~\onlinecite{our1} is applied. As is seen the
$\Phi$-versus-$S$ curve shows more rapid than $S^{-1}$ decreasing
in wide area range for both cases considered.

\begin{figure}
\epsfclipon
 \epsfxsize=\linewidth
 \epsfbox{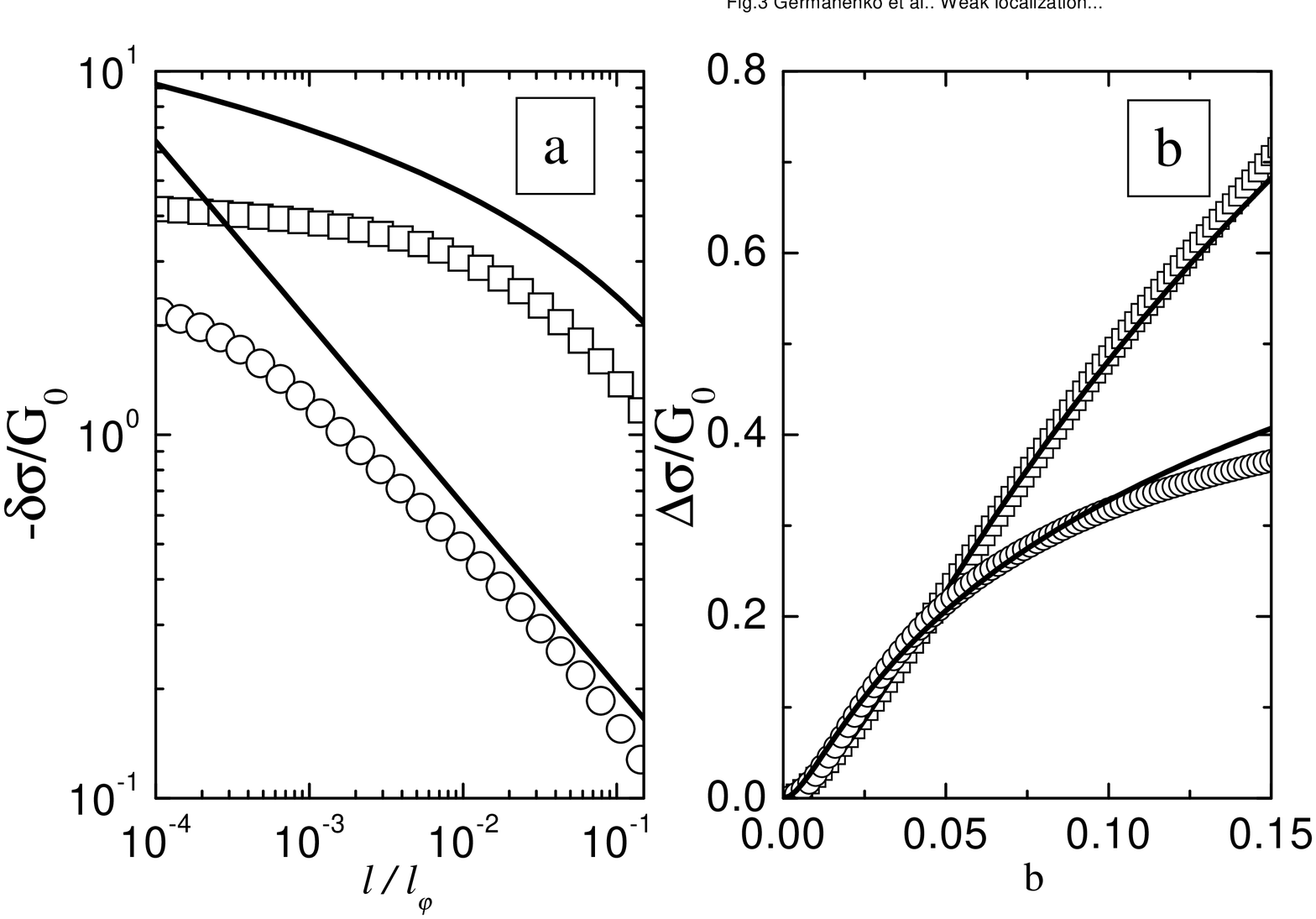}
 \caption
{(a) The interference quantum correction to the conductivity in
zero magnetic field as a function of $l/l_\varphi$ obtained from
the simulation (symbols), calculated from Eq. (\ref{eq20}) (upper
curve) and Eq. (\ref{eq21}) (lower curve). (b) The magnetic field
dependence of interference quantum correction to the conductivity
obtained from the simulation  with $l/l_\varphi=3.5\times 10^{-3}$
(symbols). Curves are the best fit to Eq.\ (\ref{eq3}), carried
out for $b\le 0.1$. Symbolic designations are the same as in
Fig.~\ref{fig2}. } \label{fig3}
\end{figure}

We turn now to consideration of the interference quantum
correction to the conductivity. It has been found using
Eq.~(\ref{eq09}) and discrete form of Eq.\ (\ref{eq1}) [see Ref.\
\onlinecite{our1} for details]:
\begin{equation}
{\cal W}_i=\frac{1}{I_s d\, l}\sum_{k}
\cos\left(\frac{bS_k}{l^2}\right) \exp \left(-
\frac{l_k}{l_\varphi} \right),\; i=c,p. \label{eq2}
\end{equation}
Here, summation runs over all closed paths, $S_k$, $l_k$ are the
area enclosed and the length of $k$th closed path, respectively,
$b$ is magnetic field measured in units of $B_{tr}$. Deriving
Eq.~(\ref{eq2}) we have supposed that $\tau_\varphi$ is the same
within the channel and puddles. In this case ${\cal W}_i$ are
independent of the Fermi velocity $v_F$. Figure\ \ref{fig3}(a)
shows the simulation results for $\delta\sigma(b=0)$, obtained
for different $l/l_\varphi$ values, and, for comparison, the
results of theoretical calculation obtained through the well-known
formulas\cite{c1,satT5}
\begin{equation}
 \delta\sigma=-G_0 \ln\left(1+\frac{l_\varphi}{l}\right)
 \label{eq20}
\end{equation}
for 2D case, and
\begin{equation}
 \delta\sigma=-6.094\, G_0 L_c^{-1} \sqrt{l_\varphi l}
 \label{eq21}
\end{equation}
for quasi one-dimensional case with $L_c$ as channel length. As is
seen the simulation results reveal the behavior close to
theoretical ones only for considerably large $l$ to $l_\varphi$
ratios: $l/l_\varphi>0.01$. Experimentally, this corresponds to
the high temperatures. With lowering $l/l_\varphi$, i.e., with
temperature decrease, the simulation data tend to saturate as
opposed to the theories. It is clear that such a behavior is a
result of steep fall in $\Phi(S)$ dependencies at large areas
(see Fig.~\ref{fig3}(b)).

Let us analyze the magnetic field dependencies of interference
quantum correction. As an example, in Fig.~\ref{fig3}(b) we
present the dependencies
$\Delta\sigma(b)=\delta\sigma(b)-\delta\sigma(0)$ calculated with
$l/l_\varphi=3.5\times 10^{-3}$. We have considered these data as
experimental ones and fitted them to the well-known
expression\cite{hik,schm}
\begin{equation}
\Delta\sigma(b)=a G_0 \left[
\psi\left(\frac{1}{2}+\frac{\gamma}{b}\right)-
\psi\left(\frac{1}{2}+\frac{1}{b}\right)- \ln{\gamma} \right],
\label{eq3}
\end{equation}
where $\psi(x)$ is a digamma function. The parameters $a$ and
$\gamma=l/l_\varphi$ have been used as fitting ones. Just this
procedure is usually used to determine the value of $l_\varphi$ in
real 2D samples.  Results of the fit are represented in
Fig.~\ref{fig3}(b) by curves. As is seen, Eq.\ (\ref{eq3})
satisfactorily describes the simulation data.

\begin{figure}
 \epsfclipon
 \epsfxsize=8.3cm
 \epsfbox{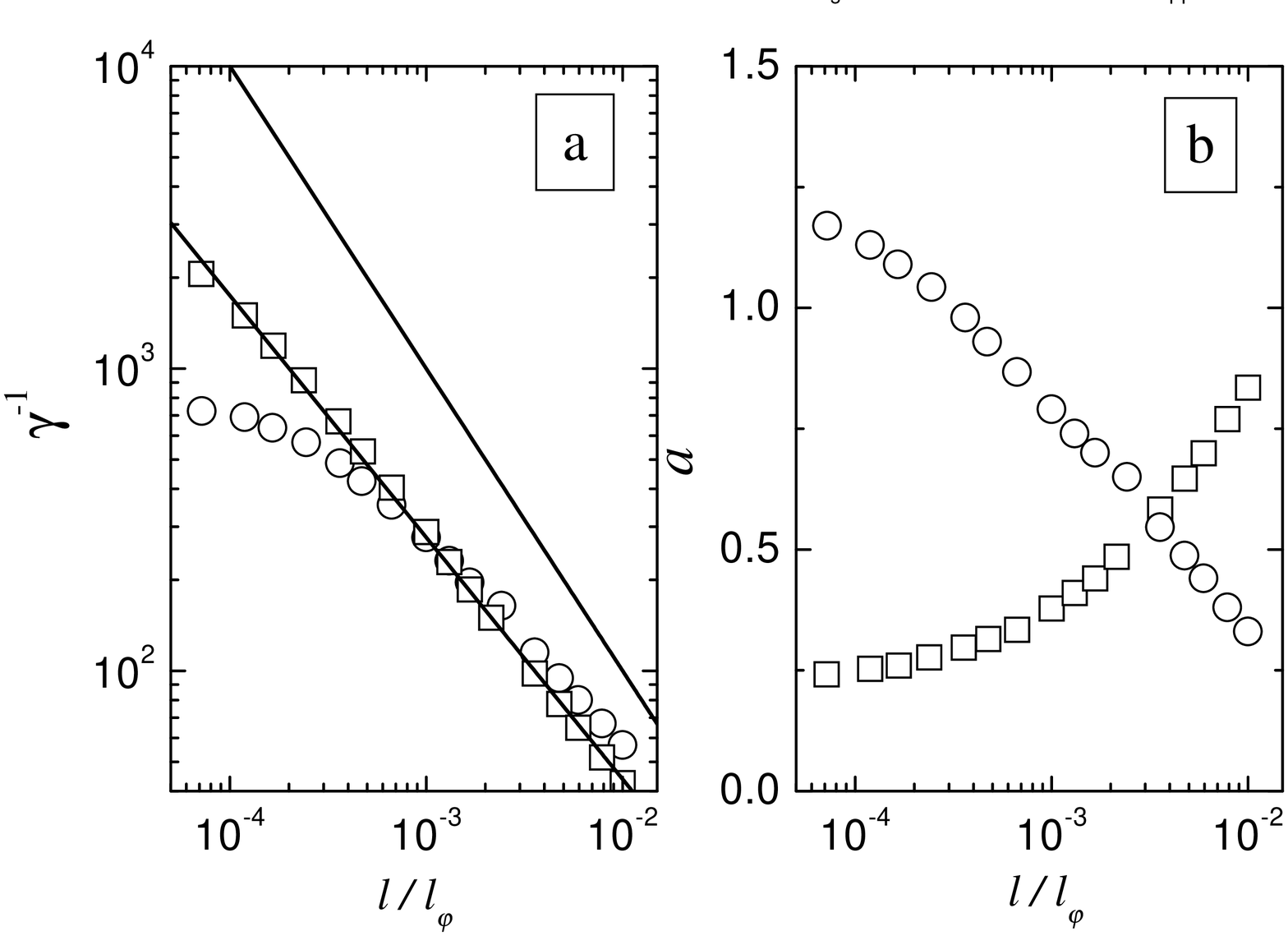}
\caption{The fitting parameters $\gamma$ (a) and $a$ (b) as
function of $l/l_\varphi$ ratio. Upper line in (a) is
$\gamma=l/l_\varphi$, lower is $\gamma \propto (l/l_\varphi)
^{0.8}$. Symbolic designations are the same as in
Fig.~\ref{fig2}.}
 \label{fig4}
\end{figure}

How the value of $\gamma$, found from the fitting, matches the
value of $l/l_\varphi=\tau/\tau_\varphi$, put in Eq.\ (\ref{eq2}),
is shown in Fig.~\ref{fig4}(a). The fitting procedure gives the
value of $\gamma$ which is less than the value of $l/l_\varphi$.
In the case of square channel, $l/l_\varphi$-dependence of
$\gamma$ can be described by the power function: $\gamma\propto
(l/l_\varphi)^{0.8}$. As for the case of long channel, the
$\gamma$-versus-$l/l_\varphi$ plot shows saturation at
$l/l_\varphi<10^{-3}$. On the assumption of $l_\varphi\propto
T^{-\alpha}$, $\alpha>0$, Figure\ \ref{fig4}(a) shows qualitative
temperature dependencies of the ``phase breaking time" as it is
obtained from the standard fitting procedure. Thus, the presence
of macroscopical inhomogeneity in samples can lead to
inconsistency between temperature dependence of the ``phase
breaking time", found experimentally, and the real temperature
dependence of $\tau_\varphi$. This inconsistency may be
qualitative for inhomogeneous samples in which the conduction
through the long channels dominates: the value obtained from the
fitting procedure can be saturated, whereas the real phase
breaking time decreases with decrease of temperature and diverges
at $T\to 0$. Using quota we underline that the ``phase breaking
time"  is nothing more than the fitting parameter here. It should
not be always identified with the real value of phase breaking
time.

Finally, we present the $l/l_\varphi$-dependence of prefactor $a$
(Fig.~\ref{fig4}(b)). Absolute value of $a$ has no meaning,
because its value should depend on size distribution of channels
in real sample. However, the dependence is meaningful. In both
cases considered the value of prefactor $a$ strongly depends on
$l$ to $l_\varphi$ ratio, and the direction of change of $a$ with
growing $l/l_\varphi$ is determined by the kind of inhomogeneity:
for long channel, the value of $a$ increases with decrease of
$l/l_\varphi$, whereas it decreases for the case of square
channel.

Note, that the results presented cannot be extrapolated down to
$l/l_\varphi=0$ [or $T=0$]. The point is that for
$l/l_\varphi<2\times 10^{-5}$ the value of
$L_\varphi=\sqrt{l_\varphi l}$ becomes greater than the size of
the element, which determines the correlation length $\xi_p$. In
this case our system  should behave itself as homogeneous one.
Thus, it seems reasonable to say that the $l/l_\varphi$-dependence
of $\gamma$ must tend to $\gamma\propto l_\varphi^{-1}$, and the
value of prefactor must be equal to unity for $L_\varphi\gg
\xi_p\simeq 10^4$.

In conclusion, we have numerically studied the statistics of
closed paths and the negative magnetoresistance in macroscopically
inhomogeneous 2D systems. We have considered the situation when
the main contribution to the quantum interference correction to
the conductivity comes from constrictions. Two types of
constrictions have been considered: long and square channels. We
have shown that the magnetic field dependence of the negative
magnetoresistance is well fitted to the Hikami expression for both
systems. However, the value of the fitting parameter  $\gamma$
does not coincide with the value $l/l_\varphi$ fed into
simulation, and, moreover, the $\gamma$-versus-$l/l_\varphi$ plot
is not linear. This means that the temperature dependence of the
``phase breaking time", as it is obtained experimentally from the
standard analysis of the negative magnetoresistance for
inhomogeneous 2D system, can differ drastically from the real
temperature dependence of $\tau_\varphi$. An indicator, whether
the system is inhomogeneous, is the strong temperature dependence
of the prefactor. The Fourier transformation of the magnetic field
dependencies of the negative magnetoresistance taken at different
temperatures\cite{our1,our2,our3} can be used in order to obtain
the function $\Phi(S)$, which carries direct information about
the area distribution function of closed paths, and, consequently,
about the character of inhomogeneity of the samples, investigated.

The results presented can be directly used to interpret the
weak-localization experiments carried out on the artificial
diffusive cavities and constrictions. To our knowledge there is
deficit of the paper devoted to such objects. The main attention
was attracted to the ballistic cavities and systems with antidot
arrays [see, for example, Ref.~\onlinecite{cav}].

The authors are grateful to I.~V.~Gornyi for many valuable
discussions. This work was supported in part by the RFBR through
Grants No. 00-02-16215 and No. 01-02-17003, the Program {\it
University of Russia} through Grants No. 990409 and No. 990425,
and the CRDF through Award No. REC-005.


\begin{thebibliography}{}
\bibitem{c1} B. L. Altshuler, and A. G. Aronov, in {\em
Electron-Electron Interaction in Disordered Systems}, edited by
A.~L.~Efros and M.~Pollak, (North Holland, Amsterdam, 1985) p.1.

\bibitem{c2}
P. A. Lee and T. V. Ramakrishnan, Rev. Mod. Phys., {\bf 57}, 287
(1985), G. Bergman, Physics Reports, {\bf 107}, 1 (1984), B. L.
Altshuler, A. G. Aronov, D. E. Khmelnitski, and A. I. Larkin, in
{\em Quantum theory of solids}, edited by I.~M.~Lifshitz, (Mir
Publishers, Moscow, 1982).

\bibitem{satT1} P. Mohanty, E. M. Q. Jarivala, and R. A. Webb,
Phys. Rev. Lett. {\bf 78}, 3366 (1997), P. Mohanty and R. A. Webb,
Phys. Rev. B {\bf55}, 13452 (1997), C.~Prasad, D.~K.~Ferry,
A.~Shailos, M.~Elhassen, J.~P.~Bird, L.-H.~Lin, N.~Aoki,
Y.~Ochiai, K.~Ishibashi, and Y.~Aoyagi, Phys. Rev. B {\bf62},
15356 (2000).

\bibitem{satT22}
M.~E.~Gershenson, Annalen der Physik {\bf 8}, 559 (1999).

\bibitem{satT3}
B.~L.~Altshuler, M.~E.~Gershenson, and I.~L.~Aleiner, Physica E
{\bf 3}, 58 (1998), B.~L.~Altshuler, I.~L.~Aleiner, and
M.~E.~Gershenson. Phys. Rev. Lett. {\bf 82}, 3190 (1999).

\bibitem{satT5}
I.~L.~Aleiner, B.~L.~Altshuler, and M.~E.~Gershenson. Waves in
Random Media {\bf 9}, 201 (1999).

\bibitem{satT6} Dmitrii S.~Golubev, and Andrei D.~Zaikin,
Phys. Rev. B {\bf 59}, 9195 (1999), Phys. Rev. B {\bf 62}, 14061
(2000).

\bibitem{hik} S.~Hikami, A.~Larkin and Y.~Nagaoka,
Prog. Theor. Phys. {\bf 63}, 707 (1980).

\bibitem{schm} H.-P.~Wittman and A.~Schmid, J. Low. Temp. Phys.
{\bf 69}, 131 (1987).

\bibitem{our1}G.~M.~Minkov, A.~V.~Germanenko, O.~E.~Rut, and
I.~V.~Gornyi, Phys. Rev. B {\bf 61}, 13164 (2000).

\bibitem{chak} S.~Chakravarty and A.~Schmid, Phys. Reports {\bf 140},
193 (1986).

\bibitem{dyak} M.~I.~Dyakonov,  Solid St. Comm. {\bf 92}, 711 (1994).

\bibitem{dmit} I. V. Gornyi, A. P. Dmitriev, and V. Yu. Kachorovskii,
Phys. Rev. B {\bf 56}, 9910 (1997).

\bibitem{our2}G.~M.~Minkov, S.~A.~Negashev, O.~E.~Rut,
A. V. Germanenko, O. I. Khrykin, V.~I.~Shashkin, and
V.~M.~Danil'tsev,  Phys. Rev. B {\bf 61}, 13172 (2000).

\bibitem{our3}G.~M.~Minkov, A.~V.~Germanenko, O.~E.~Rut,
O. I. Khrykin, V.~I.~Shashkin and V.~M.~Danil'tsev, Phys. Rev. B
{\bf 62}, 17089 (2000).


\bibitem{AronAED}
A.~G.~Aronov, M.~E.~Gershenzon, and Yu.~E.~Zhuravlev, Zh. Eksp.
Teor. Fiz., {\bf 87} 971 (1984) [Sov. Phys. JETP {\bf 60}, 554
(1984)].

\bibitem{Pal} A.~Palevskiand and G.~Deutscher, Phys. Rev. B {\bf 34},
431 (1986).

\bibitem{Dump} G.~Dumpich and A.~Carl, Phys. Rev. B, {\bf 43}
12074 (1991).

\bibitem{fried}S. Friedrichowski, A. Carl, and G. Dumpich,
Europhys. Lett., {\bf 32}(3), 247 (1995).

\bibitem{Stauf}
D.~Staufer, {\em Introduction to Percolation Theory} (Taylor \&
Francis, London, 1985).

\bibitem{samokh} K.~V.~Samokhin, Phys. Rev. E {\bf 59}, R2501
(1999).

\bibitem{cav} I.~L.~Aleiner and A.~I.~Larkin, Phys. Rev. B {\bf 54},
14423 (1996), Bodo Huckestein, Roland Ketzmerick, Caio H.
Lewenkopf, Phys. Rev. Letters {\bf 84}, 5504 (2000).
\end{thebibliography}
\end{document}